\documentclass[usenatbib, useAMS,referee]{biom} 
\usepackage{epsfig,epsf,psfrag}
\usepackage{lscape}

\usepackage{amsbsy,amsmath,amssymb,mathrsfs ,color}
\usepackage{verbatim}
\usepackage{booktabs, enumerate}

\usepackage{amsfonts}
\usepackage{graphicx}
\usepackage{enumerate}
\usepackage{subcaption}
\usepackage{setspace}
\usepackage{hyperref}
\usepackage{multirow}
\usepackage{pdflscape}
\usepackage{longtable}
\usepackage{lscape}
\usepackage{bm}
\usepackage{arydshln}
\usepackage{xcolor}
\usepackage{blkarray}
\usepackage{multirow}

\title[Overall marginalized models for longitudinal zero-inflated count data]
{Overall marginalized models for longitudinal zero-inflated count data}

\author{Keunbaik Lee$^{1,*}$\email{keunbaik@skku.edu},
Eun Jin Jang$^{2,**}$ 
Dipak Dey$^{3,***}$ 
\\
$^{1}$Department of Statistics, Sungkyunkwan University, Seoul, 03063, Korea \\
$^{2}$Department of Data Science, Andong National University, Andong, Gyungbuk 36729, Korea\\
$^{3}$Department of Statistics, University of Connecticut, Storrs, Connecticut 06269, USA}

\begin{document}

\label{firstpage}


\begin{abstract}
To analyze longitudinal zero-inflated count data, we extend existing models by introducing marginalized zero-inflated Poisson (MZIP) models with random effects, which explicitly capture the marginal effect of covariates and address limitations of previous methods. These models provide a clearer interpretation of the overall mean effect of covariates on zero-inflated count data. To further accommodate overdispersion, we develop marginalized zero-inflated negative binomial (MZINB) models. Both models incorporate subject-specific heterogeneity through a flexible random effects covariance structure. Simulation studies are conducted to
evaluate the performance of the MZIP and MZINB models, comparing their inference under both homogeneous and heterogeneous random effects. Finally, we illustrate the applicability of the proposed models through an analysis of systemic lupus erythematosus data.
\end{abstract}

%



\maketitle

\section{Introduction}\label{sect:1}

Longitudinal count data with an excess of zeros is a common challenge in fields like ecology, actuarial science, public health, and medicine. This is particularly evident in medical studies, especially when focusing on rare events such as hospitalizations. Our paper specifically examines hospitalization counts for patients with systemic lupus erythematosus (SLE), which represent this type of data.

To analyze count data with many zeros, the zero-inflated Poisson (ZIP) model is a standard approach \cite{Mullahy:1986}. This model cleverly combines two components: a binary one to account for the presence of zeros, and a Poisson one for the observed counts. Within this framework, zeros can arise from a ``structural'' process (generating only zeros) or the standard Poisson counting process (which can also produce zeros). \shortcite{Lambert:1992} further developed the ZIP regression model for overdispersed data, where a logistic part models the probability of a zero count and a Poisson part models the count frequency, conditional on non-zero outcomes.

Standard ZIP models typically use two sets of regression parameters: one for the Poisson mean and another for the probability of being an excess zero, which can be interpreted as latent classes. However, it is crucial to recognize that the Poisson mean in these models is conditional, not representative of the overall mean. To address this, \shortcite{Long:etal:2014} developed the marginalized ZIP model to capture the overall mean effect of covariates. This was later extended by \shortcite{Preisser:etal:2016} to zero-inflated negative binomial data. Our work extends \shortcite{Long:etal:2014}'s marginalized models  to handle longitudinal zero-inflated count data.

Various zero-inflated Poisson (ZIP) models with random effects have been developed to analyze longitudinal zero-inflated count data, primarily by capturing within-subject dependence in the Poisson component \cite{Hall:2000}. For example, \shortcite{Min:Agesti:2005} proposed two-part zero-inflated random effects models for clustered count data. \shortcite{Lee:etal:2006} then incorporated shared subject-specific random effects into both components of the zero-inflated model to account for zero inflation and over-dispersion in longitudinal counts. \shortcite{Lee:etal:2011} further extended these by proposing marginalized ZIP and marginalized zero-inflated negative binomial (ZINB) models. More recently, \shortcite{Zhu:etal:2017} developed ZIP and ZINB random effects models with heterogeneous random effects covariance matrices.

Despite these advances, none of these models explicitly address the overall mean effect of covariates. While \shortcite{Long:etal:2015} did propose a marginalized ZIP model with random effects for longitudinal zero-inflated count data, their approach conditions the overall mean effect of covariates on the random effects. This makes it unsuitable for truly capturing marginal effects.

Our work is driven by the analysis of systemic lupus erythematosus (SLE) data. SLE is a chronic, complex autoimmune disease where the immune system attacks the body's own tissues. This leads to widespread inflammation and organ damage, affecting various organs such as the skin, joints, kidneys, lungs, brain, and blood vessels. The disease's progressive nature, coupled with its diverse and changing symptoms, creates significant challenges for diagnosis and long-term management, often resulting in frequent hospitalizations and elevated healthcare costs \cite{Kan:etal:2013, Murimi-Worstell:2021}.

The SLE patient hospitalization data we analyze is longitudinal zero-inflated count data, characterized by a large proportion of zeros. Since patients with SLE show distinct differences in characteristics between those frequently hospitalized and those who are not \cite{Lee:etal:2016}, a specific analytical approach is crucial. Therefore, instead of separately identifying factors associated with the probability of having more than zero hospitalizations and factors influencing the number of hospitalizations among those with hospitalizations, we focus on estimating the marginal effect. This approach captures the average effect across all SLE patients, including those with zero hospitalizations.

Existing models have often limitations when analyzing longitudinal zero-inflated count data with its unique real-world characteristics. To address this, we propose novel models specifically designed to capture the marginal effects of covariates on the overall mean response. Our approach effectively incorporates random effects to account for between-subject variations and can also analyze nonzero overdispersed data. A key innovation is our ability to not only model covariate effects on the overall mean response—a feature absent in typical zero-inflated models—but also to allow the random effects covariance matrix to vary with covariates, enabling heterogeneous covariance. This flexible structure offers significant advantages by accurately capturing both correlation and subject-level heterogeneity. As a result, our models are particularly well-suited for complex datasets common in clinical and epidemiological research, especially where between-subject variations are heterogeneous.

The rest of the paper is structured as follows. Section 2 introduces a marginalized ZIP model with random effects for longitudinal zero-inflated count data, emphasizing the marginal effect of covariates and presenting the maximum likelihood estimation approach. Additionally, we propose a marginalized ZINB model with random effects to address zero-inflated overdispersed longitudinal count data. Section 3 provides simulation results demonstrating the performance and comparison of both models under both homogeneous and heterogeneous random effects. In Section 4, we apply these models to analyze systemic lupus erythematosus data. Finally, Section 5 summarizes our key findings.

\section{Proposed models}

In this section, we propose two marginalized zero-inflated models  for longitudinal zero-inflated count data to represent the marginal effect of the covariates. Let $Y_{it}$, $i=1,\ldots, N$ and $t=1,\ldots, n_i$, be a count variable that takes the value zero from a Bernoulli distribution. Let $x_{it}$ be covariates corresponding to $Y_{it}$.

\subsection{Marginalized ZIP random effects model}\label{Sec:MZIPREM}

We first develop a marginalized ZIP random effects model (MZIPREM) to account for the marginal effect of covariates. MZIPREM is a model that explains the marginal effects of covariates on the overall mean of the outcome by transforming \cite{Lee:etal:2011}'s model. This model also explains both the Poisson mean and the probability of an excess zero in the typical ZIP model.

The MZIPREM consists of three submodels: the marginal model, the dependence model, and the overall mean model. The first two models are from \cite{Lee:etal:2011}. The marginal model is a standard marginal ZIP model, while the dependence model explains the correlation in repeated count data. Finally, the overall mean model demonstrates the effect of the covariates on the overall mean. The marginal model is given by:\\
\noindent\textbf{Marginal model: }\\
\begin{eqnarray}\label{model:marginal}
P(y_{it} | x_{it})=
    \left\{
      \begin{array}{ll}
        p_{it}^M+\left(1-p_{it}^M\right)e^{-\lambda_{it}^M}, & \hbox{for $y_{it}=0$;} \\
        \left(1-p_{it}^M\right)\frac{e^{-\lambda_{it}^M}\left(\lambda_{it}^M\right)^{y_{it}}}{y_{it}!}, & \hbox{for $y_{it}=1,2,\ldots$.}
      \end{array}
    \right.
\end{eqnarray}
with commonly used linear models
\begin{eqnarray}
    &&\mbox{logit}\left(p_{it}^M\right)=x_{it1}^T\gamma, \label{marginal-1}\\
    &&\log\left(\lambda_{it}^M\right)=x_{it2}^T\beta+\log(\mbox{off}_{it}), \label{marginal-2}
\end{eqnarray}
where $p_{it}^M$ is mixing proportion,
$\gamma$ is $p_1\times 1$ parameter vector associated with the excess zeros, $\beta$ is $p_2\times 1$ parameter vector associated with Poison distribution, and $x_{it1}$ and $x_{it2}$ are covariate vectors for excess zero and Poisson distribution, respectively. Thus, $\gamma$ and $\beta$ allow subgroup interpretations, providing inferences on the likelihood of belonging to the excess zero class and the mean of the at-risk class (non-excess zero group), respectively. The offset term $\text{off}_{it}$ accounts for differential exposure times.

To explain serial correlation of repeated responses, we need the dependence model with random effects. The dependence model is given by:\\
\noindent\textbf{Dependence model: }
\begin{eqnarray}\label{model:dependence}
P(y_{it}| b_i)=
    \left\{
      \begin{array}{ll}
        p_{it}^c(b_i)+\left(1-p_{it}^c(b_i)\right)e^{-\lambda_{it}^c(b_i)}, & \hbox{for $y_{it}=0$;} \\
        \left(1-p_{it}^c(b_i)\right)\frac{e^{-\lambda_{it}^c(b_i)}\left(\lambda_{it}^c(b_i)\right)^{y_{it}}}{y_{it}!}, & \hbox{for $y_{it}=1,2,\ldots$.}
      \end{array}
    \right.
\end{eqnarray}
with
\begin{eqnarray}
    &&\Phi^{-1}\left(p_{it}^c(b_i)\right)=\Delta_{it1}+z_{it1}^Tb_{i1}, \label{Depend-1}\\
    &&\log\left(\lambda_{it}^c(b_i)\right)=\Delta_{it2}+z_{it2}^Tb_{i2}, \label{Depend-2}\\
    &&b_i=(b_{i1}^T,b_{i2}^T)^T \stackrel{iid}{\sim}  N(0,\Sigma_i), \label{Depend-3}
\end{eqnarray}
where $z_{it1}$ and $z_{it2}$ are respectively $q_1\times 1$ and $q_2\times 1$ covariate vectors, $b_{i1}$ and $b_{i2}$ are corresponding random effects vectors, and
$$\Sigma_i=\left(
         \begin{array}{cc}
            \Sigma_{i1} & \Sigma_{i,12} \\
            \Sigma_{i,12}^T & \Sigma_{i2} \\
         \end{array}
       \right),$$
with $\Sigma_i$ being a subject-specific positive-definite covariance matrix. The submatrices $\Sigma_{11}$ and $\Sigma_{22}$ represent the covariance structures of the random effects associated with the structural zero component and the Poisson count component, respectively. The off-diagonal submatrix $\Sigma_{12}$ captures the correlation between the random effects influencing the occurrence of structural zeros and the Poisson counts.

From (\ref{marginal-1}), (\ref{marginal-2}), (\ref{Depend-1}), and (\ref{Depend-2}), we have the following relationships:
\begin{eqnarray}
    &&p_{it}^M=E\left(p_{it}^c(b_i)\right), \nonumber\\
    &&\lambda_{it}^M=E\left(\lambda_{it}^c(b_i)\right).\label{relation}
\end{eqnarray}
Thus, $\Delta_{it1}$ and $\Delta_{it2}$ are deterministic parameters given $\gamma$, $\beta$, and $\Sigma$, and they are given by
\begin{eqnarray}
 &&\Delta_{it1}=\sqrt{1+z_{it1}^T\Sigma_1z_{it1}}\Phi^{-1}\left(p_{it}^M\right), \label{delta-1}\\
  &&\Delta_{it2}=x_{it2}^T\beta-\frac{1}{2}z_{it2}^T\Sigma_2z_{it2}+\log\mbox{off}_{it}. \label{delta-2}
\end{eqnarray}

Note that the model (\ref{marginal-2}) is specifically for the Poisson process and does not directly address the overall population mean of the outcome. Consider the marginal mean of $Y_{it}$, denoted as $\mu_{it}=E(Y_{it})$, which is often the primary interest of researchers. From (\ref{relation}), the marginal mean $\mu_{it}$ is given by:
\begin{eqnarray}\label{marginal-relation}
\mu_{it}=\left(1-p_{it}^M\right)\lambda_{it}^M.
\end{eqnarray}
In (\ref{marginal-relation}), the overall population mean ($\mu_{it}$) is a function of all covariates and parameters from both the marginal and dependence models. For the $j$th covariate ($x_{it2j}$) in $x_{it2}$ where $x_{it1}=x_{it2}$, the ratio of means for a one-unit increase in $x_{it2j}$ is:
\begin{eqnarray*}
\frac{E(Y_{ij}|x_{it2j}+1,\tilde{x}_{it2})}{E(Y_{ij}|x_{it2j},\tilde{x}_{it2})}=\exp(\beta_{j})\frac{1+\exp\left(x_{it2j}\gamma_j+\tilde{x}_{it2}^T\tilde{\gamma}\right)}{1+\exp\left((x_{it2j}+1)\gamma_j+\tilde{x}_{it2}^T\tilde{\gamma}\right)},
\end{eqnarray*}
where $\tilde{x}_{it2}$ represents all covariates except $x_{it2j}$, and $\tilde{\gamma}$ is obtained by excluding $\gamma_j$ from $\gamma$. Therefore, unless $\gamma_j=0$, the incidence density ratio (IDR) varies across different levels of the covariates included in the logistic part of the marginal model.

{
The model proposed by \cite{Lee:etal:2011} is limited because it does not directly show the effect of covariates on the overall mean ($\mu_{it}$), as evidenced by the IDR. Furthermore, assessing the variability of any IDR estimates at fixed levels of non-exposure covariates requires formal statistical techniques, such as the delta method or bootstrap resampling. To address this, we introduce a new model that can directly represent covariate effects on the population mean.}

We define the marginal mean $\mu_{it}$ as:\\
\noindent\textbf{Overall mean model:}
\begin{eqnarray}\label{model-marginal}
\log \mu_{it} = x_{it2}^T \alpha+\log(\mbox{off}_{it}),
\end{eqnarray}
where $\alpha$ is a $p_1\times 1$ vector. Our proposed overall marginalized models average across the two ZIP model processes to yield overall effect estimates for expected counts, offering parameter interpretations consistent with a Poisson regression model.

{
To use the ZIP model likelihood framework, we first redefine as $\lambda_{it}^M=\exp\left(\eta_{it}^M\right)$, where $\eta_{it}^M$ does not necessarily have to be a linear function of the model parameters \cite{Long:etal:2014}. Instead, solving the equation (\ref{marginal-relation}) and substituting equations (\ref{marginal-1}) and (\ref{model-marginal}) yields the following expression for $\eta_{it}^M$:
\begin{eqnarray}\label{marginal-relation-1}
\eta_{it}^M=x_{it2}^T\alpha+\log\left(1+e^{x_{it1}^T\gamma}\right).
\end{eqnarray}
}

As the simple random intercept form of models is often adequate in practice, we only discuss the case with $b_{i1}$ and $b_{i2}$ being univariate with $z_{it1}=z_{it2}=1$. Then we have the decomposition
\begin{eqnarray*}
    \Sigma_i=\left(
         \begin{array}{cc}
            \sigma_{i1}^2 & \rho_i\sigma_{i1}\sigma_{i2} \\
            \rho_i\sigma_{i1}\sigma_{i2} & \sigma_{i2}^2 \\
         \end{array}
       \right)
       =D_iR_iD_i,
\end{eqnarray*}
where
\begin{eqnarray*}
  D_i=\left(\begin{array}{cc}
                 \sigma_{i1} & 0 \\
                 0 & \sigma_{i2} \\
              \end{array}\right),~~ R_i=\left(\begin{array}{cc}
                 1 & \rho_i \\
                 \rho_i & 1 \\
              \end{array}
        \right).
\end{eqnarray*}

Since the random effect covariance matrix $\Sigma_i$ can be heteroskedastic, we assume generalized linear models for $\sigma_{i1}$, $\sigma_{i2}$, and $\rho_{i}$ as follows:
\begin{eqnarray}
    &&\log\sigma_{i1}=h_i^T\zeta_1, \label{sigma-1}\\
    &&\log\sigma_{i2}=h_i^T\zeta_2, \label{sigma-2}\\
    &&\frac{1}{2}\log\left(\frac{1+\rho_i}{1-\rho_i}\right)=w_i^T\delta,\label{sigma-3}
\end{eqnarray}
where $\zeta_1$, $\zeta_2$, and $\delta$ are respectively unknown parameters of dimensions $a\times 1$, $b\times 1$, and $c\times 1$, and $h_i$ and $w_i$ are subject-specific covariate vectors.

Note that the two models for $\sigma_{i1}$ and $\sigma_{i2}$ are log-linear models, ensuring that the estimates of $\sigma_{i1}$ and $\sigma_{i2}$ are positive. Additionally, the Fisher Z-transformation of $\rho_i$ guarantees that the range of $\rho_i$ is between $-1$ and $1$. As a result, the estimation of $\Sigma_i$ is guaranteed to be positive definite and heteroskedastic.

\subsection{Marginalized ZINB random effects model}

To analyze nonzero overdispersed data, we propose the marginalized zero-inflated negative binomial random effects model (MZINBREM). The negative binomial distribution is suitable for overdispersed data, and the zero-inflated negative binomial (ZINB) model offers greater flexibility than the zero-inflated Poisson (ZIP) model. Specifically, ZINB handles overdispersion arising from both excess zeros and heterogeneity in the count component, while ZIP only addresses overdispersion due to excess zeros.

The MZINBREM also consists of the three submodels presented in Subsection \ref{Sec:MZIPREM}:
First the marginal model is given by:\\
\noindent\textbf{Marginal model: }
\begin{eqnarray}\label{MZINBREM-1}
   P(y_{it} | x_{it})=\left\{
      \begin{array}{ll}
        p_{it}^M+\left(1-p_{it}^M\right)\left(1+\nu\lambda_{it}^M\right)^{-\frac{1}{\nu}}, & \hbox{for $y_{it}=0$;} \\
        \left(1-p_{it}^M\right)\frac{\Gamma\left(\nu^{-1}+y_{it}\right)}{\Gamma\left(\nu^{-1}\right)y_{it}!}\left(1+\nu\lambda_{it}^M\right)^{-\frac{1}{\nu}}\left(\frac{\nu\lambda_{it}^M}{1+\nu\lambda_{it}^M}\right)^{y_{it}}, & \hbox{for $y_{it}=1,2,\ldots$,}
      \end{array}
    \right.
\end{eqnarray}
using the same models for $p_{it}^M$ and $\lambda_{it}^M$ as defined in (\ref{marginal-1}) and (\ref{marginal-2}).

The dependence model is as follows:\\
\noindent\textbf{Dependence model: }
\begin{eqnarray}\label{MZINBREM-2}
    P(y_{it} | b_i)=\left\{
      \begin{array}{ll}
        p_{it}^c(b_{i})+\left(1-p_{it}^c(b_{i})\right)\left(1+\nu\lambda_{it}^c(b_i)\right)^{-\frac{1}{\nu}}, & \hbox{for $y_{it}=0$;} \\
        \left(1-p_{it}^c(b_i)\right)\frac{\Gamma\left(\nu^{-1}+y_{it}\right)}{\Gamma\left(\nu^{-1}\right)y_{it}!}\left(1+\nu\lambda_{it}^c(b_i)\right)^{-\frac{1}{\nu}}\left(\frac{\nu\lambda_{it}^c(b_i)}{1+\nu\lambda_{it}^c(b_i)}\right)^{y_{it}}, & \hbox{for $y_{it}=1,2,\ldots$,}
      \end{array}
    \right.
\end{eqnarray}
using the same models for $p_{it}^c(b_{i})$ and $\lambda_{it}^c(b_i)$ as defined in (\ref{Depend-1})-(\ref{Depend-3}).

Similar to (\ref{relation}), we have deterministic parameters $\Delta_{it1}$ and $\Delta_{it2}$, which take the same forms as in (\ref{delta-1}) and (\ref{delta-2}). Additionally, we define the overall mean model, which follows the same structure as in (\ref{model-marginal}). The models for $\Sigma_i$ are identical to those in (\ref{sigma-1})–(\ref{sigma-3}).

\section{Estimation}

\subsection{Estimation in MZIPREM}

We begin by outlining the parameter estimation process for the MZIPREM model. Let $\theta = (\gamma^\top, \alpha^\top, \zeta_1^\top, \zeta_2^\top, \delta^\top)^\top$. The likelihood function for the MZIPREM is then given by:
\begin{eqnarray*}
  L(\theta; y)&=&\prod_{i=1}^N \int L(\theta, b_i;y_i)\phi(b_i)db_i,
\end{eqnarray*}
where
\begin{eqnarray*}
  L(\theta, b_i;y_i)&=&\prod_{t=1}^{n_i}\left\{p_{it}^c(b_i)+\left(1-p_{it}^c(b_i)\right)e^{-\lambda_{it}^c(b_i)}\right\}^{I_{(y_{it}=0)}}\left\{\left(1-p_{it}^c(b_i)\right)\frac{e^{-\lambda_{it}^c(b_i)}\left(\lambda_{it}^c(b_i)\right)^{y_{it}}}{y_{it}!}\right\}^{1-I_{(y_{it}=0)}},
\end{eqnarray*}
and $\phi(\cdot)$ is the bivariate normal density function for $N(0,\Sigma_i)$ which is reexpressed by
\begin{eqnarray*}
  \phi(b_i)=(2\pi)^{-1}\exp\Biggl\{-\frac{1}{2}\Biggl(h_i^T\zeta_1+h_i^T\zeta_2\Biggr)-\frac{1}{2}\log\Biggl(1-\rho_i^2\Biggr)-\frac{1}{2}b_i^T\Sigma_i^{-1}b_i\Biggr\}
\end{eqnarray*},
and $I_A$ is the indicator function.

Note that $\Delta_{it1}$ and $\Delta_{it2}$ in (\ref{delta-1}) and (\ref{delta-2}) are also reparameterized as follow:
\begin{eqnarray}
    &&\Delta_{it1}=\sqrt{1+\sigma_{i1}^2} \Phi^{-1}\left(p_{it}^M\right),\\
    &&\Delta_{it2}=\log \lambda_{it}^M-\frac{1}{2}\sigma_{i2}^2.
\end{eqnarray}

The log likelihood function is given by
\begin{eqnarray}
 \log L(\theta; y)&=&\sum_{i=1}^N\log L(\theta; y_i)=\sum_{i=1}^N \log \int L(\theta, b_i ; y_i)\phi(b_i)db_i \nonumber\\
 &=&\sum_{i=1}^N\log\int\exp \Biggl[\sum_{t=1}^{n_i}\biggl\{ \log\left(1-p_{it}^c(b_i)\right)+I_{(y_{it}=0)}\log\left(\frac{p_{it}^c(b_i)}{1-p_{it}^c(b_i)}+e^{-\lambda_{it}^c(b_i)}\right) \nonumber\\
 &&+\left(1-I_{(y_{it}=0)}\right)\biggl(-\lambda_{it}^c(b_i)+y_{it}(\Delta_{it2}+b_{i2})-\log y_{it}!\biggr) \biggr\} \Biggr]\phi(b_i)d b_i.\label{like-1}
\end{eqnarray}
Since the integrals in (\ref{like-1}) are not in closed form, we use the Gauss-Hermite quadrature \cite{Anderson:Aitkin:1985} to numerically integrate out the random effects.

Maximizing the log likelihood with respect to $\theta$ yields the likelihood equations:
{\scriptsize
\begin{eqnarray}
&&\frac{\partial\log L(\theta;y)}{\partial\gamma}=\sum_{i=1}^N\frac{1}{L(\theta; y_i)}\int L(\theta, b_i; y_i)\sum_{t=1}^{n_i}\Biggl\{\frac{I_{(y_{it}=0)}}{1+\frac{1-p_{it}^c(b_i)}{p_{it}^c(b_i)}e^{-\lambda_{it}^c(b_i)}}-p_{it}^c(b_i)\Biggr\}\frac{\phi(\Delta_{it1}+z_{it1}^Tb_{i1})}{p_{it}^c(b_i)\left(1-p_{it}^c(b_i)\right)}\frac{\partial\Delta_{it1}}{\partial\gamma}\phi(b_i)d b_i, \label{score-1}\\
&&\frac{\partial\log L(\theta;y)}{\partial\alpha}=\sum_{i=1}^N\frac{1}{L(\theta; y_i)}\int L(\theta, b_i; y_i)\sum_{t=1}^{n_i}\Biggl\{-\frac{I_{(y_{it}=0)}\lambda_{it}^c(b_i)e^{-\lambda_{it}^c(b_i)}}{\frac{p_{it}^c(b_i)}{p_{it}^c(b_i)\left(1-p_{it}^c(b_i)\right)}+e^{-\lambda_{it}^c(b_i)}}+\left(1-I_{(y_{it}=0)}\right)\left(y_{it}-\lambda_{it}^c(b_i)\right)\Biggr\}
\frac{\partial\Delta_{it2}}{\partial\alpha}\phi(b_i)db_i, \nonumber\\
 \label{score-2}\\
&&\frac{\partial\log L(\theta;y)}{\partial \zeta_{1l}}=\sum_{i=1}^N\frac{1}{L(\theta; y_i)}\int
L(\theta, b_i; y_i)\Biggl[\sum_{t=1}^{n_i}\Biggl\{I_{(y_{it}=0)}-\left(p_{it}^c(b_i)+(1-p_{it}^c(b_i)e^{-\lambda_{it}^c(b_i)})\right)\Biggr\} \nonumber\\ &&~~~~~~~~~~~~~~~~~~~~~~~~~~~~~~~~~~~\times\frac{\phi(\Delta_{it1}+b_{11})}{(1-p_{it}^c(b_i))\left(p_{it}^c(b_i)+(1-p_{it}^c(b_i)e^{-\lambda_{it}^c(b_i)})\right)}\frac{\partial\Delta_{it1}}{\partial \zeta_{il}}-\left(\frac{1}{2}h_{il}+\frac{1}{2}b_i^T\frac{\partial\Sigma_i^{-1}}{\partial\zeta_{1l}}b_i\right)\Biggr]\phi(b_i)db_i, \label{score-3}\\
&&\frac{\partial\log L(\theta;y)}{\partial \zeta_{2l}}=\sum_{i=1}^N\frac{1}{L(\theta; y_i)}\int
L(\theta, b_i; y_i)\sum_{t=1}^{n_i}\Biggl\{-I_{(y_{it}=0)}\Biggl(y_{it}-\frac{p_{it}^c(b_i)\lambda_{it}^c(b_i)}{p_{it}^c(b_i)+\left(1-p_{it}^c(b_i)\right)e^{-\lambda_{it}^c(b_i)}}\Biggr)\frac{\partial\Delta_{it2}}{\partial \zeta_{2l}}+\left(y_{it}-\lambda_{it}^c(b_i)\right)\frac{\partial\Delta_{it2}}{\partial \zeta_{2l}} \nonumber\\
&&~~~~~~~~~~~~~~~~~~~~~~~~~~~~~~~~~~~~-\left(\frac{1}{2}h_{il}+\frac{1}{2}b_i^T\frac{\partial\Sigma_i^{-1}}{\partial\zeta_{2l}}b_i\right)\Biggr\}\phi(b_i)db_i, \label{score-4}\\
&&\frac{\partial\log L(\theta;y)}{\partial \delta_l}=\sum_{i=1}^N\frac{1}{L(\theta; y_i)}\int
L(\theta, b_i; y_i)\Biggl(\frac{\rho_i}{1-\rho_i^2}\frac{\partial\rho_i}{\partial\delta_l}-\frac{1}{2}b_i^T\frac{\partial\Sigma_i^{-1}}{\partial\delta_l}b_i\Biggr)\phi(b_i)db_i, \label{score-5}
\end{eqnarray}
}
where
\begin{eqnarray*}
&&L(\theta ; y_i)=\int L(\theta, b_i;y_i)\phi(b_i)db_i,\\
&&\frac{\partial\Delta_{it1}}{\partial\gamma}=\sqrt{1+\sigma_{i1}^2}\frac{p_{it}^M\left(1-p_{it}^M\right)x_{it1}}{\phi\left(\Phi^{-1}(p_{it}^M)\right)},~~~\frac{\partial\Delta_{it2}}{\partial\alpha}=x_{it2},\\
&&\frac{\partial\Delta_{it1}}{\partial \zeta_{1l}}=\frac{\sigma_{i1}^2}{\sqrt{1+\sigma_{i1}^2}}h_{il}\Phi^{-1}(p_{it}^M),~~~\frac{\partial\Delta_{it2}}{\partial\zeta_{2l}}=-\sigma_{i2}^2h_{il},\\
&&\frac{\partial\Sigma_i^{-1}}{\partial\zeta_{1l}}=
\frac{\partial D_i^{-1}}{\partial\zeta_{1l}}R_i^{-1}D_i^{-1}+D_i^{-1}R_i^{-1}\frac{\partial D_i^{-1}}{\partial\zeta_{1l}},\\
&&\frac{\partial\Sigma_i^{-1}}{\partial\zeta_{2l}}=
\frac{\partial D_i^{-1}}{\partial\zeta_{2l}}R_i^{-1}D_i^{-1}+D_i^{-1}R_i^{-1}\frac{\partial D_i^{-1}}{\partial\zeta_{2l}},\\
&&\frac{\partial\rho_i}{\partial\delta_l}=4w_{il}\frac{\exp\left(2w_i^T\delta\right)}{\left\{1+\exp\left(2w_i^T\delta\right)\right\}^2},\\
&&\frac{\partial\Sigma_i^{-1}}{\partial\delta_l}=D_i^{-1}\frac{\partial R_i^{-1}}{\partial\delta_l}D_i^{-1}
\end{eqnarray*}
with
\begin{eqnarray*}
&&\frac{\partial D_i^{-1}}{\zeta_{1l}}=\left(\begin{array}{cc}
                 -h_{il}/\sigma_{i1} & 0 \\
                 0 & 0 \\
              \end{array}\right),~~~
\frac{\partial D_i^{-1}}{\zeta_{2l}}=\left(\begin{array}{cc}
                 0 & 0 \\
                 0 & -h_{il}/\sigma_{i2} \\
              \end{array}\right),\\
&&\frac{\partial R_i^{-1}}{\partial \delta_l}=\left(\begin{array}{cc}
                 \frac{2\rho_i}{(1-\rho_i^2)^2}\frac{\partial\rho_i}{\partial\delta_l} & -\frac{1+\rho_i^2}{(1-rho_i^2)^2}\frac{\partial\rho_i}{\partial\delta_l} \\
                 -\frac{1+\rho_i^2}{(1-rho_i^2)^2}\frac{\partial\rho_i}{\partial\delta_l} & \frac{2\rho_i}{(1-\rho_i^2)^2}\frac{\partial\rho_i}{\partial\delta_l} \\
              \end{array}\right).
\end{eqnarray*}

The matrix of second derivatives of the observed data log likelihood is quite complex. However, in any correctly specified model, the sample empirical covariance matrix of the individual scores acts as a consistent estimator of the information matrix, involving only the first derivatives. Therefore, the Quasi-Newton method can be applied to solve the likelihood equations using the update:
\begin{eqnarray*}
\theta^{(g+1)} = \theta^{(g)} + \left[ H(\theta; y)^{-1} \frac{\partial \log L(\theta; y)}{\partial \theta} \right]_{\theta = \theta^{(g)}},
\end{eqnarray*}
where $H(\theta; y)$ is an empirical and consistent estimator of the information matrix at step $g$, given by
\begin{eqnarray*}
H(\theta; y)=\sum_{i=1}^N\frac{\partial \log L(\theta; y_i)}{\partial\theta} \frac{\partial \log L(\theta; y_i)}{\partial\theta^T}.
\end{eqnarray*}

{
At convergence, the large-sample variance-covariance matrix of the parameter estimates is then obtained as the inverse of $H(\hat{\theta}; y)$.
}

\subsection{Estimation in MZINBREM}

Let $\psi=(\gamma^T, \alpha^T, \zeta_1^T, \zeta_2^T, \delta^T, \nu)^T$. Similarly, the likelihood function for the MZINBREM is given by:
\begin{eqnarray}
   L(\psi; y)&=&\prod_{i=1}^N\int L(\psi, b_i;y_i)\phi(b_i) db_i,\label{like-NB}
\end{eqnarray}
where
\begin{eqnarray*}
 L(\psi, b_i;y_i)&=&\prod_{t=1}^{n_i}\left\{p_{it}^c(b_i)+\left(1-p_{it}^c(b_i)\right)\left(1+\nu\lambda_{it}^c(b_i)\right)^{-\frac{1}{\nu}}\right\}^{I_{(y_{it}=0)}}\\
 ~~~&&\left\{\left(1-p_{it}^c(b_i)\right)\frac{\Gamma\left(\nu^{-1}+y_{it}\right)}{\Gamma\left(\nu^{-1}\right)y_{it}!}\left(1+\nu\lambda_{it}^c(b_i)\right)^{-\frac{1}{\nu}}\left(\frac{\nu\lambda_{it}^c(b_i)}{1+\nu\lambda_{it}^c(b_i)}\right)^{y_{it}}\right\}^{1-I_{(y_{it}=0)}}.
\end{eqnarray*}
The log likelihood function is given by:
{\small
\begin{eqnarray*}
&&\log L(\psi; y)=\sum_{i=1}^N\log\int\exp\left[\sum_{t=1}^{n_i}\left\{I_{(y_{it}=0)}\log\left(p_{it}^c(b_i)+\left(1-p_{it}^c(b_i)\right)\left(1+\nu\lambda_{it}^c(b_i)\right)^{-\frac{1}{\nu}}\right)\right.\right.\\
&&\left.\left.+\left(1-I_{(y_{it}=0)}\right)\left(\log\left(1-p_{it}^c(b_i)\right)+\sum_{l=0}^{y_{it}-1}\log\left(1+\nu l\right)-\left(y_{it}+\frac{1}{\nu}\right)\log\left(1+\nu\lambda_{it}^c(b_i)\right)+y_{it}\log\lambda_{it}^c(b_i)-\log\left(y_{it}!\right)\right)\right\}\right]\phi(b_i)db_i.
\end{eqnarray*}
}
The integral in (\ref{like-NB}) is calculated numerically using the Gauss-Hermite quadrature. Maximizing the log likelihood with respect to $\psi$ yields the likelihood equations:
{\small
\begin{eqnarray}
  &&\frac{\partial\log L(\psi; y)}{\partial\gamma}=\sum_{i=1}^N\frac{1}{L(\psi; y_i)}\int L(\psi, b_i; y_i)\left\{\frac{I_{(y_{it}=0)}}{1+\frac{1-p_{it}^c(b_i)}{p_{it}^c(b_i)}\left(1+\nu\lambda_{it}^c(b_i)\right)^{-\frac{1}{\nu}}}-p_{it}^c(b_i)\right\}\frac{\phi\left(\Delta_{it1}+z_{it1}^Tb_i\right)}{P_{it1}^c(b_i)\left(1-P_{it1}^c(b_i)\right)}\frac{\partial\Delta_{it1}}{\partial\gamma}\phi(b_i)db_i,\\
  &&\frac{\partial\log L(\psi; y)}{\partial\alpha}=\sum_{i=1}^N\frac{1}{L(\psi; y_i)}\int L(\psi, b_i; y_i)\sum_{t=1}^{n_i}\left\{-I_{(y_{it}=0)}\frac{\lambda_{it}^c(b_i)\left(1+\nu\lambda_{it}^c(b_i)\right)^{-\frac{1}{\nu}-1}}{\frac{p_{it}^c(b_i)}{1-p_{it}^c(b_i)}+\left(1+\nu\lambda_{it}^c(b_i)\right)^{-\frac{1}{\nu}}}\frac{\partial\Delta_{it2}}{\partial\alpha}\right. \nonumber\\
  &&~~~~~~~~~~~~~~~~~~~~~~~~~~~\left.\left(1-I_{(y_{it}=0)}\right)\left(y_{it}-\lambda_{it}^c(b_i)\right)\frac{1}{1+\nu\lambda_{it}^c(b_i)}\frac{\partial\Delta_{it2}}{\partial\alpha}\right\}\phi(b_i)db_i,\\
  &&\frac{\partial\log L(\psi; y)}{\partial\zeta_{1l}}=\sum_{i=1}^N\frac{1}{L(\psi; y_i)}\int L(\psi, b_i; y_i)\left[\sum_{t=1}^{n_i}\left\{\frac{I_{(y_{it}=0)}}{1+\frac{1-p_{it}^c(b_i)}{p_{it}^c(b_i)}\left(1+\nu\lambda_{it}^c(b_i)\right)^{-\frac{1}{\nu}}}-p_{it}^c(b_i)\right\}\frac{\phi\left(\Delta_{it1}+z_{it1}^Tb_i\right)}{p_{it}^c(b_i)\left(1-p_{it}^c(b_i)\right)}\frac{\partial\Delta_{it1}}{\partial\zeta_{1l}}\right.\nonumber\\
  &&~~~~~~~~~~~~~~~~~~~~~~~~\left.-\left(h_{il}+\frac{1}{2}b_i^T\frac{\partial\Sigma_i^{-1}}{\partial\zeta_{1l}}b_i\right)\right]\phi(b_i)db_i,\\
  &&\frac{\partial\log L(\psi; y)}{\partial\zeta_{2l}}=\sum_{i=1}^N\frac{1}{L(\psi; y_i)}\int L(\psi, b_i; y_i)\left[\sum_{t=1}^{n_i}\left\{-I_{(y_{it}=0)}\frac{\lambda_{it}^c(b_i)\left(1+\nu\lambda_{it}^c(b_i)\right)^{-\frac{1}{\nu}-1}}{\frac{p_{it}^c(b_i)}{1-p_{it}^c(b_i)}+\left(1+\nu\lambda_{it}^c(b_i)\right)^{-\frac{1}{\nu}}}\frac{\partial\Delta_{it2}}{\partial\zeta_{2l}}\right.\right.\nonumber\\
  &&~~~~~~~~\left.\left. +\left(1-I_{(y_{it}=0)}\right)\left(y_{it}-\lambda_{it}^c(b_i)\right)\frac{1}{1+\nu\lambda_{it}^c(b_i)}\frac{\partial\Delta_{it2}}{\partial\zeta_{2l}}\right\}-\left(h_{il}+\frac{1}{2}b_i^T\frac{\partial\Sigma_i^{-1}}{\partial\zeta_{2l}}b_i\right)\right]\phi(b_i)db_i,\\
  &&\frac{\partial\log L(\psi; y)}{\partial\delta_l}=\sum_{i=1}^N\frac{1}{L(\psi; y_i)}\int L(\psi, b_i; y_i)\left(\frac{\rho_i}{1-\rho_i^2}\frac{\partial\rho_i}{\partial\delta_l}-\frac{1}{2}b_i^T\frac{\partial\Sigma_i^{-1}}{\partial\delta_l}b_i\right)\phi(b_i)db_i,\\
  &&\frac{\partial\log L(\psi; y)}{\partial\nu}=\sum_{i=1}^N\frac{1}{L(\psi; y_i)}\int L(\psi, b_i; y_i)\sum_{t=1}^{n_i}\left\{I_{(y_{it}=0)}\frac{\frac{1}{\nu}\left(\log\left(1+\nu\lambda_{it}^c(b_i)\right)-\frac{\nu\lambda_{it}^c(b_i)}{1+\nu\lambda_{it}^c(b_i)}\right)}{\frac{p_{it}^c(b_i)}{1-p_{it}^c(b_i)}\left(1+\nu\lambda_{it}^c(b_i)\right)^{\frac{1}{\nu}}+1}\right. \nonumber\\
  &&~~~~~~~\left. +\left(1-I_{(y_{it}=0)}\right)\left(\sum_{l=0}^{y_{it}-1}\frac{l}{1+\nu l}+\frac{1}{\nu^2}\log\left(1+\nu\lambda_{it}^c(b_i)\right)-\left(y_{it}+\frac{1}{\nu}\right)\frac{\lambda_{it}^c(b_i)}{1+\nu\lambda_{it}^c(b_i)} \right)\right\}\phi(b_i)db_i
\end{eqnarray}
}
The information matrix $H(\psi; y)$ is given by
\begin{eqnarray*}
H(\psi; y)=\sum_{i=1}^N\frac{\partial \log L(\psi; y_i)}{\partial\psi} \frac{\partial \log L(\psi; y_i)}{\partial\psi^T}.
\end{eqnarray*}

We also use the Quasi-Newton method to solve the likelihood equations using the update:
\begin{eqnarray*}
\psi^{(g+1)} = \psi^{(g)} + \left[ H(\psi; y)^{-1} \frac{\partial \log L(\psi; y)}{\partial \psi} \right]_{\theta = \psi^{(g)}}.
\end{eqnarray*}

\section{Simulation studies}

In this section, we conduct several simulation studies to evaluate the performance of the proposed method under two distinct scenarios. For each scenario, we generate data based on the MZIPREM and MZINBREM models, respectively. We then assess the effectiveness of the proposed models in estimating regression coefficients, providing a comprehensive evaluation of their accuracy and robustness. The studies were all conducted using R version 4.3.2 on a PC equipped with an AMD Ryzen 9 5950X 16-core processor and 128 GB of RAM.

\subsection{Scenario 1: MZIPREM}
We generated 500 random count data from a marginalized ZIP random effects model under (\ref{model:marginal}) and (\ref{model:dependence})  across different sample sizes with two covariates including group and time.

For $t=1,\ldots, n_i$, $x_{it1}^\top\gamma$, $x_{it2}^{\top}\alpha$, $h_i^{\top}\zeta_1$, $h_i^{\top}\zeta_2$, and $w_i^{\top}\delta$ are given by
\begin{eqnarray}
  &&x_{it1}^{\top}\gamma=\gamma_0+\gamma_1 group_i+\gamma_2 Time_{it},~~x_{it2}^{\top}\alpha=\alpha_0+\alpha_1 group_i+\alpha_2 Time_{it}, \label{sim-1}\\
  &&h_i^{\top}\zeta_1=\zeta_{10},~~h_i^{\top}\zeta_2=\zeta_{20},~~w_i^{\top}\delta=\delta_0, \label{sim-2}
\end{eqnarray}
where $\gamma=(-2.8, 0.58,  0.1)^{\top}$, $\alpha=(1.6, -0.4,  0.1)^{\top}$, $\zeta_{10}=-0.1$, $\zeta_{20}=-0.1$, and $\delta_0=0.8$.

To evaluate the estimation accuracy of the proposed model, we calculated the percent relative bias (PRB), mean of standard errors (SE), standard deviation (SD), coverage probability (CP), and the mean of the $M$ estimated mean parameters (MEAN) using the following formulas:
\begin{eqnarray*}
  &&MEAN(\xi)=\frac{1}{M}\sum_{m=1}^M\hat{\xi}_m\stackrel{def}{=}\bar{\hat{\xi}},~~PRB(\xi)=\frac{\bar{\hat{\xi}}-\xi}{\xi}\times 100,\\
  &&SE(\xi)=\frac{1}{M}\sum_{m=1}^M se(\hat{\xi}_m),~~SD(\xi)=Stdev\left(\hat{\xi}_1,\ldots,\hat{\xi}_m\right),\\
  &&CP(\xi)=\frac{1}{M}\sum_{m=1}^M I_{\left(\hat{\xi}_m\pm 1.96 se(\hat{\xi}_m)\right)}(\xi),
\end{eqnarray*}
where $\xi$ is one of the parameters $\theta$, $se(\xi_m)$ is the standard error of $\xi$ in dataset $m$, $I_{(A)}(\xi)=1$ for $\xi\in A$ and 0 otherwise, and $M=500$.

We also present the average of absolute PRBs (APRB), SEs (ASE), SDs (ASD), and coverage probabilities (ACP) for the parameters, $\gamma$ and $\alpha$, which are calculated as follows:
{\small
\begin{eqnarray*}
  APRB=\frac{1}{6}\sum_{l=1}^{6} \lvert PRB(\xi_l) \rvert,~~~
  ASE=\frac{1}{6}\sum_{l=1}^{6} SE(\xi_l),~~
  ASD=\frac{1}{6}\sum_{l=1}^{6} SD(\xi_l),~~~
  ACP=\frac{1}{6}\sum_{l=1}^{6} CP(\xi_l).
\end{eqnarray*}
} Additionally, to evaluate the estimation accuracy of the random effects covariance matrix $\Sigma_i$, we present the Frobenius norm (FROB), which is computed as:
\begin{eqnarray*}
    tr\left(\hat{\Sigma_i}\Sigma_i^{-1}-I\right)^2.
\end{eqnarray*}

The simulation results in Table \ref{table_Simu-1} illustrate the performance of the MLEs for the parameters across different sample sizes ($N=300$, 700, and 1000). As the sample size increases, the average percentage relative bias (APRB), standard errors (ASE), and standard deviations (ASD) decrease, indicating improved estimation accuracy. Across all sample sizes, the coverage probabilities remain close to the nominal 95\% level (ACP = 95.4\% for $N=300$, 95.6\% for $N=700$, and 94.4\% for $N=1000$). Furthermore, the table shows that ASE and ASD are approximately equal for all estimated parameters, suggesting consistent estimation performance. Additionally, the Frobenius norms decrease as the sample size increases from $N=300$ to $N=1000$, demonstrating that the estimated covariance matrix converges to the true covariance matrix with larger sample sizes.

\begin{table}[h]
  \caption{Simulation study results obtained from generated samples using MZIP model. The mean of the estimate (MEAN), percent relative bias (PRB), coverage probability (CP), average standard error (SE), standard deviation (SD) of 500 estimates, Frobeniusnorm (FROB), average absolute value of the PRBs (APRB), average values of the SEs and SDs are displayed.}
  \begin{center}
    \begin{tabular}{lcrcrcr}
      \hline
         & \multicolumn{2}{c}{$N=300$} & \multicolumn{2}{c}{$N=700$} & \multicolumn{2}{c}{$N=1000$} \\
      \cline{2-7}
                   & Mean    & PRB  &  Mean   & PRB  &  Mean & PRB  \\
                   & $\mbox{SE}_{(\mbox{SD})}$ & CP   & $\mbox{SE}_{(\mbox{SD})}$ & CP   & $\mbox{SE}_{(\mbox{SD})}$ & CP \\
      \hline
       $\gamma_0$  &  -2.835           &  1.26 & -2.810            &  0.34 &   -2.802     &  0.07     \\
        (-2.8)     & $0.371_{(0.395)}$ & 93.0  & $0.231_{(0.233)}$ & 95.2  & $0.190_{(0.195)}$ &  94.4    \\
       $\gamma_1$  &   0.595           &  2.66 &  0.594            &  2.49 &   0.569      & -1.88    \\
        ( 0.58)    & $0.362_{(0.368)}$ &  96.6 & $0.229_{(0.231)}$ & 95.8  & $0.191_{(0.194)}$ &  95.6    \\
       $\gamma_2$  &  0.124            & 23.87 &  0.093            & -7.47 &   0.114      & 13.90     \\
        ( 0.1)     & $0.461_{(0.448)}$ &  96.8 & $0.291_{(0.296)}$ & 95.8  & $0.237_{(0.238)}$ &  94.2    \\
       \hline
       $\alpha_0$  & 1.592             & -0.47 &  1.594            & -0.35 &    1.597      & -0.21     \\
        ( 1.6)     & $0.093_{(0.090)}$ & 95.4  & $0.058_{(0.061)}$ & 93.4  & $0.048_{(0.049)}$ &  93.6    \\
       $\alpha_1$  & -0.391            & -2.22 & -0.394            & -1.44 &   -0.398      & -0.50    \\
        (-0.4)     & $0.116_{(0.110)}$ &  95.0 & $0.074_{(0.075)}$ & 94.8  & $0.061_{(0.064)}$ &  93.8    \\
       $\alpha_2$  & 0.098             & -2.47 &  0.101            &  0.59 &    0.101          &  0.99    \\
        ( 0.1)     & $0.065_{(0.067)}$ &  94.2 & $0.041_{(0.040)}$ & 95.6  & $0.033_{(0.034)}$ &  93.6    \\
      \hline
       $1000\times FROB(\hat{\Sigma})$  & \multicolumn{2}{c}{7.57} & \multicolumn{2}{c}{6.04}  &   \multicolumn{2}{c}{2.47}  \\
       APRB        & \multicolumn{2}{c}{5.49} & \multicolumn{2}{c}{4.32}   & \multicolumn{2}{c}{2.92}\\
       $\mbox{ASE}_{\mbox{(ASD)}}$ & \multicolumn{2}{c}{$0.228_{(0.229)}$} & \multicolumn{2}{c}{$0.143_{(0.143)}$} & \multicolumn{2}{c}{$0.117_{(0.119)}$}\\
       ACP         & \multicolumn{2}{c}{95.4}  & \multicolumn{2}{c}{95.6}   & \multicolumn{2}{c}{94.4}\\
      \hline
    \end{tabular}\label{table_Simu-1}
  \end{center}
\end{table}

\subsection{Scenario 2: MZINBREM}

We generated 500 random count datasets from a marginalized ZINB random effects model based on equations (\ref{MZINBREM-1}) and (\ref{MZINBREM-2}) across three sample sizes of $N=300$, $700$, and $1000$. The parameter values were set as in (\ref{sim-1}) and (\ref{sim-2}), with $\nu=0.8$.

Table \ref{table_Simu-2} summarizes the simulation results, reporting PRBs, SEs, SDs, CPs, APRBs, ASEs, ACPs, and the Frobenius norms of $\Sigma$. Consistent with the results from the MZIPREM simulation study, APRBs, ASEs, and ASDs decrease as the sample size increases, indicating good performance of the estimation method. Additionally, the Frobenius norms decrease with increasing sample size—from $N=300$ to $N=1000$—suggesting that the estimated covariance matrix approaches the true covariance structure with larger samples.

Overall, these results provide strong evidence that the estimation procedures for both MZIPREM and MZINBREM yield accurate and reliable parameter estimates across the simulation settings.

\begin{table}[h]
  \caption{Simulation study results obtained from generated samples using MZINB model. The mean of the estimate (MEAN), percent relative bias (PRB), coverage probability (CP), average standard error (SE), standard deviation (SD) of 500 estimates, Frobeniusnorm (FROB), average absolute value of the PRBs (APRB), average values of the SEs and SDs are displayed.}
  \begin{center}
    \begin{tabular}{lcrcrcr}
      \hline
         & \multicolumn{2}{c}{$N=300$} & \multicolumn{2}{c}{$N=700$} & \multicolumn{2}{c}{$N=1000$} \\
      \cline{2-7}
                   & Mean    & PRB  &  Mean   & PRB  &  Mean & PRB  \\
                   & $\mbox{SE}_{(\mbox{SD})}$ & CP   & $\mbox{SE}_{(\mbox{SD})}$ & CP   & $\mbox{SE}_{(\mbox{SD})}$ & CP \\
      \hline
       $\gamma_0$  &  -2.814             &  0.50   &  -2.848           &  1.70   &  -2.827           &  0.97   \\
        (-2.8)     &  $0.765_{(0.735)}$  &  93.0   & $0.493_{(0.466)}$ &  96.2   & $0.421_{(0.373)}$ & 96.2    \\
       $\gamma_1$  &   0.560             & -3.39   &   0.607           &  4.59   &   0.582           &  0.28     \\
        ( 0.58)    &  $0.566_{(0.596)}$  &  98.2   & $0.352_{(0.363)}$ &  96.0   & $0.295_{(0.294)}$ & 96.6    \\
       $\gamma_2$  &   0.145             &  45.45  &   0.140           &  40.48  &   0.130           & 29.66   \\
        ( 0.1)     &  $0.738_{(0.866)}$  &  96.0   & $0.458_{(0.482)}$ &  96.4   & $0.391_{(0.376)}$ & 97.2     \\
       \hline
       $\alpha_0$  &   1.591             & -0.54   &   1.595           & -0.29  &   1.605           &   0.30     \\
        ( 1.6)     &   $0.136_{(0.127)}$ &  95.8   & $0.087_{(0.081)}$ &  98.0   & $0.073_{(0.066)}$ & 96.0     \\
       $\alpha_1$  &  -0.398             & -0.38   &  -0.408           &  1.96   &  -0.401           &  0.17    \\
        (-0.4)     &   $0.145_{(0.124)}$ &  97.0   & $0.094_{(0.085)}$ &  96.6   & $0.079_{(0.068)}$ &  98.2    \\
       $\alpha_2$  &   0.097             & -3.21   &   0.101           &  1.10   &    0.093          & -7.08     \\
        ( 0.1)     &   $0.157_{(0.142)}$ &  96.6   & $0.101_{(0.092)}$ &  97.0   & $0.086_{(0.076)}$ &  96.0    \\
       $\nu$       &   0.792             & -1.05   &   0.795           & -0.57   &    0.797          &  -0.38    \\
        ( 0.8)     &   $0.127_{(0.102)}$ &  94.6   & $0.084_{(0.066)}$ &  97.2   & $0.070_{(0.059)}$ &  96.0    \\
      \hline
       $1000\times FROB(\hat{\Sigma})$  & \multicolumn{2}{c}{12.45} & \multicolumn{2}{c}{8.41}  &   \multicolumn{2}{c}{7.17}  \\
       APRB        & \multicolumn{2}{c}{10.70}     & \multicolumn{2}{c}{7.02}   & \multicolumn{2}{c}{6.41}\\
       $\mbox{ASE}_{\mbox{(ASD)}}$  & \multicolumn{2}{c}{$0.407_{(0.362)}$}   & \multicolumn{2}{c}{$0.257_{(0.233)}$} & \multicolumn{2}{c}{$0.204_{(0.186)}$}\\
       ACP         & \multicolumn{2}{c}{96.4}     & \multicolumn{2}{c}{97.1}   & \multicolumn{2}{c}{96.6}\\

      \hline
    \end{tabular}\label{table_Simu-2}
  \end{center}
\end{table}

\subsection{Comparsion of heteroskedastic and homogeneous $\Sigma_i$}

We also conducted an additional simulation study to compare the performance of the proposed models under both homogeneous and heterogeneous random effects covariances. Specifically, we generated 500 datasets from the MZIP and MZINB models incorporating heterogeneous random effects covariance as follows:
\begin{eqnarray*}
  h_i^T\xi_1=\xi_{10}+group_i\xi_{11},~~h_i^T\xi_2=\xi_{20}+group_i\xi_{21},
\end{eqnarray*}
where $(\xi_{10}, \xi_{11}, \xi_{20}, \xi_{21})=(-0.1,0.1,-0.1,0.1)$.

We then fitted both MZIP and MZINB models assuming either heterogeneous or homogeneous random effects covariance covariances. To evaluate and compare the models, we calculated the average absolute percentage relative biases (APRB) and the average standard errors (SEs) for the parameters $\gamma$ and $\alpha$. The results are summarized in Table \ref{table-comparsion}.

The results indicate that the MZIP model with a heterogeneous random effects covariance produces a smaller APRB than the model with a homogeneous covariance, while the ASEs remain similar in both cases. A similar pattern is observed for the MZINB model, where the APRB is lower under the heterogeneous random effects covariance, and the ASEs are nearly identical to those in the MZIP model. These results demonstrate that accounting for heterogeneity in the random effects covariance improves estimation accuracy in the MZIP and MZINB models.

\begin{table}[h]
  \caption{Simulation study results obtained from generated samples using MZIP and MZNB models with heterogenenous random effects covariance matrices. Average absolute value of the PRBs (APRB), average values of the SEs are displayed.}
  \begin{center}
    \begin{tabular}{lccccc}
      \hline
       & \multicolumn{2}{c}{MZIP} & & \multicolumn{2}{c}{MZINB} \\
       \cline{2-3} \cline{5-6}
       & Homogeneous & Heterogeneous & & Homogeneous & Heterogeneous \\
      \hline
       APRB & 5.35  & 2.05  & & 8.05  & 5.11  \\
       ASE  & 0.081 & 0.085 & & 0.199 & 0.206   \\
       \hline
    \end{tabular}\label{table-comparsion}
  \end{center}
\end{table}

\section{Analysis of lupus data}\label{example}

We analyze the SLE data, as described in Section \ref{sect:1}, to better understand healthcare utilization patterns and the disease's impact on patients. For this purpose, we leverage data from the Korean National Health Insurance (NHI) system. As a government-run, nationwide program covering about 98\% of Koreans, the NHI provides comprehensive medical coverage. Its vast database contains rich, longitudinal information, including diagnoses, treatments, surgeries, prescriptions, and lab/imaging results, making it an invaluable asset for population health research \cite{Park:Lee:2021}.

Previous research by \shortcite{Kim:etal:2023} used NHI data to estimate the direct healthcare costs for individuals newly diagnosed with SLE between 2008 and 2018. Their cost estimates included a wide range of services like medications, outpatient and inpatient care, and diagnostic tests. Subsequent studies by \shortcite{Jang:etal:2025} and \shortcite{Lee:etal:2025} expanded on this by using multivariate linear models to jointly analyze three aspects of direct healthcare costs, further illuminating the disease's economic impact.

This paper focuses on modeling the number of hospitalizations over time as a longitudinal outcome to identify key factors linked to hospitalization rates in patients with severe SLE. The dataset covers a 10-year follow-up period, providing insights into trends over time and individual patient experiences. Notably, 61.7\% of observations show no hospitalizations, indicating an excess of zeros and highlighting the need for models suitable for zero-inflated count data.

The main explanatory variables examined in this analysis are gender (0 for male, 1 for female), age, and the Charlson Comorbidity Index (CCI), a common measure of overall health burden. To address skewed data and improve model stability, age and CCI were transformed to $\log(Age)$ and $CCI/10$, respectively. The dataset also contains instances of patients dropping out and missing data over time. In this study, missing data are assumed to be missing at random (MAR), allowing for valid conclusions under appropriate modeling techniques.

\subsection{Model fit}

We investigated MZIPREM and MZINBREM with different random effects covariance structures determined by subject $i$'s Gender and Age at time $t$. Table \ref{table_model-1} outlines six models. Models 1 and 2 are MZIPREMs with random effects covariance matrices dependent on Gender alone and on both Gender and Age, respectively. Models 3 and 4 are MZINBREMs with random effects covariance matrices structured similarly. Models 5 and 6 correspond to ZIP and ZINB models with random effects, respectively.

\begin{table}[h]
  \caption{The models for $h_i^T\zeta_1$,  $h_i^T\zeta_2$ and $w_i^T\delta$ are presented for the SLE data.}
  \vskip-0.3cm \hrule
    \centering
		\begin{tabular}{crll}
			\multicolumn{2}{c}{Model description} & $h_i^T\zeta_1$ and $h_i^T\zeta_2$ & $w_i^T\delta$  \\
			\hline
			Model 1  &  MZIPREM    & $\zeta_{j0}+\zeta_{j1}\mbox{gender}_i$ & $\delta_0$  \\
			Model 2  &  MZIPREM    & $\zeta_{j0}+\zeta_{j1}\mbox{gender}_i$ & $\delta_0+\delta_1\log(\mbox{Age})_i$  \\
			Model 3  &  MZINBREM   & $\zeta_{j0}+\zeta_{j1}\mbox{gender}_i$ & $\delta_0$  \\	
			Model 4  &  MZINBREM   & $\zeta_{j0}+\zeta_{j1}\mbox{gender}_i$ & $\delta_0+\delta_1\log(\mbox{Age})_i$  \\	
            Model 5  &  ZIPREM     & $\sigma_1$ and $\sigma_2$  & $\rho$ \\
            Model 6  &  ZINBREM    & $\sigma_1$ and $\sigma_2$  & $\rho$ \\
	\end{tabular}\label{table_model-1}
\hrule
\end{table}

Table \ref{table_loglik} presents the maximum likelihood estimates and Akaike Information Criterion (AIC) values. Given that Models 1 and 2, as well as Models 3 and 4, are nested, likelihood ratio tests (LRTs) were conducted. The results indicate that Model 2 provides a better fit than Model 1 (LRT $= 75.958$, p-value $< 0.0001$), and Model 4 outperforms Model 3 (LRT $= 61.874$, p-value $< 0.0001$). Since Models 2, 4, 5, and 6 are not nested, their performance was compared using AIC values. Model 4 demonstrated the lowest AIC, indicating it provides the best fit among the six models.

\begin{table}[h]
  \caption{Maximized log likelihoods and AICs for models}
  \vskip-0.3cm \hrule
    \centering
	\begin{tabular}{ccccccc}
      Model & 1 & 2 & 3 & 4 & 5 & 6 \\
      \hline
      Max.loglik & -25115.404 & -25077.430 & -24259.092 & -24228.155 & -25112.890 &  -24268.090 \\
      AIC        & 50256.808  &  50182.860  & 48546.184 & 48486.310  & 50247.780 & 48560.170 \\
      \hline
	\end{tabular}\label{table_loglik}
\hrule
\end{table}

The MLEs of the parameters for Models 1, 2, 3 and 4 are presented in Table \ref{table_MLE}. The MLEs of Models 1 and 2, and MLEs of Models 3 and 4 were similar, respectively. Since the best model in Table \ref{table_MLE} was Model 4, we focus on the results of the model. The fitted models using Model 4 were given by the following equations:
\begin{eqnarray*}
   &&\mbox{logit}\left(\hat{p}_{it}^M\right)=-3.101^*+0.663^* \log\left(\mbox{Age}_{it}\right)+0.358^* \mbox{Gender}_i-0.957^*\mbox{CCI/10}_{it},\\
   &&\log\hat{\mu}_{it}=1.751^*-0.456^*\log\left(\mbox{Age}_{it}\right)-0.223^*\mbox{Gender}_i+1.232^*\mbox{CCI/10}_{it}.
\end{eqnarray*}
The estimated probability of non-hospitalization increased with the subject's age, decreased with the subject's CCI, and was higher for females than males.
Conversely, the estimated log overall mean decreased with the subject's age, increased with the subject's CCI, and was lower for females than males.

The estimated random effects standard deviations for the probit model were $\hat{\sigma}_{i1}=\exp(-0.056 - 0.280)=0.715$ for females and $\hat{\sigma}_{i1}=\exp(-0.056)=0.946$ for males. Similarly, for the log-linear model, the estimated random effects standard deviations were $\hat{\sigma}_{i2}=\exp(-0.118 - 0.123)=0.786$ for females and $\hat{\sigma}_{i2}=\exp(-0.118)=0.889$ for males. The estimated random effects correlation for a subject with the mean $\overline{\log\mbox{Age}} = 3.637$ was
$$\hat{\rho}_i= \left\{\exp\left(2(2.994-0.818~\overline{\log\mbox{Age}})\right)-1\right\}/\left\{\exp\left(2(2.994-0.818~\overline{\log\mbox{Age}})\right)+1\right\}=0.019.$$
The corresponding random effects covariance matrices for females and males with $\overline{\log\mbox{Age}} = 3.637$ were as follows:
\begin{eqnarray*}
  \hat{\Sigma}_{\mbox{females}}= \left(
     \begin{array}{cc}
       0.511 & 0.011 \\
       0.011 & 0.618 \\
     \end{array}
  \right),~~~
  \hat{\Sigma}_{\mbox{males}}= \left(
     \begin{array}{cc}
       0.895 & 0.016 \\
       0.016 & 0.790 \\
\end{array}
  \right).
\end{eqnarray*}
These results indicate that the estimated random effects covariances differ between males and females.

\begin{table}[h]
  \caption{MLE of the parameters for Models 1-4. $^*$ indicates significance with 95\% confidence level.}
  \begin{center}
    \begin{tabular}{lrrrr}
      \hline
         & Model 1 & Model 2 & Model 3 & Model 4 \\
      \hline
      \multicolumn{2}{l}{Zero model} & & & \\
       Int. ($\gamma_0$)    & $-1.641^*$ (0.229) & $-1.857^*$ (0.244) & $-2.806^*$ (0.360) & $-3.101^*$ (0.390) \\
       log(Age) ($\gamma_1$)& $ 0.445^*$ (0.063) & $ 0.492^*$ (0.068) & $ 0.581^*$ (0.096) & $ 0.663^*$ (0.103) \\
       Gender ($\gamma_2$)  & $ 0.239^*$ (0.069) & $ 0.256^*$ (0.072) & $ 0.374^*$ (0.109) & $ 0.358^*$ (0.112) \\
       CCI/10 ($\gamma_3$)  & $-1.019^*$ (0.143) & $-0.908^*$ (0.144) & $-0.989^*$ (0.216) & $-0.957^*$ (0.215) \\
      \hline
      \multicolumn{2}{l}{Overall model} & & & \\
       Int. ($\alpha_0$)    & $ 1.996^*$ (0.131) & $ 2.046^*$ (0.158) & $ 1.635^*$ (0.207) & $ 1.751^*$ (0.213) \\
       log(Age) ($\alpha_1$)& $-0.500^*$ (0.033) & $-0.547^*$ (0.041) & $-0.420^*$ (0.056) & $-0.456^*$ (0.057) \\
       Gender ($\alpha_2$)  & $-0.268^*$ (0.050) & $-0.200^*$ (0.054) & $-0.254^*$ (0.068) & $-0.223^*$ (0.067) \\
       CCI/10 ($\alpha_3$)  & $ 1.136^*$ (0.071) & $ 1.283^*$ (0.089) & $ 1.281^*$ (0.130) & $ 1.232^*$ (0.129) \\
       Dispersion ($\nu$)   & $-~~~~~$      & $-~~~~~$ & $ 0.733^*$ (0.047) & $ 0.717^*$ (0.046) \\
      \hline
      \multicolumn{2}{l}{Random effects covariance}\\
       Int. ($\zeta_{10}$)   & $-0.375^*$ (0.095) & $-0.368^*$ (0.102) & $-0.131  $ (0.160) & $-0.056 $ (0.157) \\
       Gender ($\zeta_{11}$) & $-0.255^*$ (0.107) & $-0.219  $ (0.112) & $-0.273  $ (0.164) & $-0.280 $ (0.160) \\
       Int. ($\zeta_{20}$)   & $-0.129^*$ (0.050) & $-0.087  $ (0.050) & $-0.144^*$ (0.068) & $-0.118 $ (0.068)\\
       Gender ($\zeta_{21}$) & $-0.056^*$ (0.054) & $-0.083  $ (0.054) & $-0.139^*$ (0.072) & $-0.123 $ (0.071) \\
       Int. ($\delta_{0}$)     & $-0.099^*$ (0.048) & $ 2.436^*$ (0.317) & $-0.105^*$ (0.081) & $2.944^*$ (0.412) \\
       log(Age) ($\delta_{1}$) & $-~~~~~$     & $-0.688^*$ (0.088) &  $-~~~~~$   & $-0.818^*$ (0.115) \\
      \hline
    \end{tabular}\label{table_MLE}
  \end{center}
\end{table}

\section{Conclusion}

In this paper, we propose two overall marginalized models, MZIP and MZINB, to analyze longitudinal zero-inflated count data. These models are constructed with three distinct and complementary components: a marginal model that captures the average response across the population, a dependence model that accounts for within-subject correlation over time, and an overall model that directly reflects the marginal effects of covariates on the observed outcomes. Both the marginal and dependence components are formulated using a mixture distribution framework, combining a point mass at zero (to model structural zeros) with either a Poisson or a negative binomial distribution (to model the count data). This mixture structure enables the models to flexibly accommodate both overdispersion and the excess zeros commonly found in longitudinal count data. We appropriately address the correlation and variances of two-part random effects in the proposed framework. This includes one random component associated with the binary (zero-inflation) process and another tied to the count process (either Poisson or negative binomial). Notably, the covariance matrix of the random effects is allowed to vary as a function of subject-specific covariates, thereby enhancing model flexibility and allowing for individualized dependence structures.
Parameter estimation is carried out using a Quasi-Newton algorithm that ensures computational efficiency and convergence stability. This estimation approach enables robust inference while effectively managing the complexity of the random effects structure and the zero-inflated nature of the data.

Simulation studies demonstrate that the biases of the parameters in the dependence and overall models are negligibly small, and the maximum likelihood estimates exhibit consistency, converging towards the true parameter values with increasing sample size. Moreover, an additional simulation study was conducted to illustrate that the assumption of a homogeneous random effects covariance matrix, when heterogeneity is present and ignored, can result in biased parameter estimates. Consequently, these results emphasize the critical importance of accurately specifying the random effects covariance structure to ensure reliable parameter estimation in our proposed models.

In the analysis of the systemic lupus erythematosus (SLE) data, the MZINB model demonstrated superior model fit when compared to the MZIP model and standard ZIP and NB random effects models (ZIPREM and ZINBREM). The parameter estimates from the preferred MZINB model indicated that the predicted probability of non-hospitalization exhibited a positive relationship with age, a negative relationship with the Charlson Comorbidity Index (CCI), and a higher value for female subjects relative to male subjects. Concurrently, the estimated log of the overall mean number of hospitalizations showed a negative association with age, a positive association with CCI, and a lower value for female subjects compared to male subjects.

Future research can build on this work by extending our proposed models to overall marginalized hurdle models, which comprise a binary component for modeling the probability of zero outcomes and a truncated count model for positive counts. Hurdle models offer a more flexible framework by accommodating both zero inflation and zero deflation \cite{Min:Agesti:2005}. In contrast, zero-inflated Poisson (ZIP) models may encounter estimation challenges when the data exhibit zero deflation. This extension is part of our ongoing work.

\section*{Acknowledgments}

This project was supported by Basic Science Research Program through the National Research Foundation of Korea (KRF) funded by the Ministry of Education, Science and Technology (NRF-2022R1A2C1002752 and RS-2024-00407300 for Keunbaik Lee).


\end{document}